\documentclass[doublecol,linenumbers]{epl2} % for 2 columns style with line numbers
\usepackage{mathtools}
\usepackage{amssymb}

\newcommand{\figdir}{Figures}

\title{Classical topological order of the  Rys F-model and its breakdown in realistic spin ice: Topological sectors of Faraday loops}
\shorttitle{Topological order of the Rys F-model and its breakdown in realistic spin ice} %Insert here a short version of the title if it exceeds 70 characters

\author{Cristiano Nisoli}
\institute{Theoretical Division, Los Alamos National Laboratory, Los Alamos, NM, 87545, USA}

\abstract{
Both the Rys F-model and antiferromagnetic square ice posses the same ordered, antiferromagnetic ground state, but the ordering transition is of second order in the latter, and of infinite order in the former. To tie this difference to topological properties and their breakdown, we introduce a Faraday line representation where loops carry the energy and magnetization of the system.  Because of the absence of monopoles in the F-model, its loops have distinct topological properties, absent in square ice, and which allow for a natural partition of its phase space into topological sectors. Then the N\'eel temperature corresponds to a transition from trivial to non-trivial topological sectors. Moreover, its zero susceptibility  below a critical field is explained  by the  homotopy invariance of its magnetization. In spin ices, instead, monopoles destroy the homotopy invariance of the magnetization, and thus erase this rich topological structure. Consequently, even trivial loops can be magnetized, and their susceptibility is never zero.}

\begin{document}

\maketitle
\section{Introduction}
In 1967, Lieb solved~\cite{lieb1967exact}  the  Rys F-model~\cite{rys1963uber}  demonstrating an antiferromagnetic transition of rather unusual features: it is of infinite order and there is an order parameter, also infinitely smooth~\cite{baxter1973spontaneous}; finally, a critical field is needed to elicit magnetization below $T_c$. 

Lieb's work predated by five years the results of Kosterlitz and Thouless (KT)~\cite{kosterlitz1973ordering}, which  tied the infinitely continuous KT transition to topological properties. But regrettably, the importance of the infinite continuity in the F-model was  apparently not recognized at first, nor was it  associated to anything topological. Lieb had employed a line representation which,
while immensely clever in allowing for an exact solution via transfer matrix, is not particularly conducive to physical intuition and does not make explicit the topological features of the system. 

Our aim is not to provide new exact solutions of the F-model. It is the opposite. We seek to make explicit the topological nature of the system by mapping it to intuitive yet rigorous ``Faraday lines'', and then use it to deduce heuristically the transition and the properties of the model. 

Our Faraday lines carry all the relevant observables: energy, magnetization, parity, and $\mathbb{Z}_2$ symmetry breaking. In the F-model they are always directed loops, thus making magnetization an homotopy invariant. This, we show, explains the critical field for susceptibility. We submit that ordering corresponds to a transition between topological sectors of trivial and non-trivial loops.
While our deductions are based on heuristic arguments, our mapping to Faraday lines and the consequent partition of the phase space in topological sectors are exact. 

Various reasons motivate our conceptualization. Firstly, we wish to  elucidate how infinitely continuum transitions are related to topological sectors in a well known system.
Secondly, vertex models enjoy wide applicability and are currently studied~\cite{Baxter1982,bogoliubov2002boundary, pannevis2012critical,zinn2000six,zinn2009six,barkema1998monte,weigel2005square,keesman2017numerical,eloranta1999diamond,van1977exactly}. 
Thirdly, and generally, one wonders if the very features that make many topological  models compelling also make them physically unrealistic.

 The F-model approximates well the low-energy physics of nanomagnetic artificial square ice~\cite{Wang2006,Nisoli2013colloquium,gilbert2016frustration} as well  as of monolayer spin ice~\cite{bovo2019phase,jaubert2017spin}. And yet, these realizations possess none of its special properties~\cite{Porro2013,Zhang2013,Libal2006,ortiz2016engineering,libal2017}. Their transitions are innocuously  second order~\cite{Wu1969,levis2013thermal,cugliandolo2017artificial} (as  recently explored experimentally~\cite{sendetskyi2019continuous}), and their susceptibility is never zero. 
Our framework explains those differences: in realistic spin ice, monopoles are sink and sources of the Faraday lines, thus destroying the topological structure.

\section{Six-vertex models and the  F-model}

A {\it six-vertex model}~\cite{Baxter1982,bogoliubov2002boundary, pannevis2012critical,zinn2000six,zinn2009six,barkema1998monte,weigel2005square,keesman2017numerical,eloranta1999diamond,van1977exactly} is a set of binary spins placed on the edges of a square lattice (Fig.~1) such that only the six ice-rule obeying vertices (two spins pointing in, two pointing out~\cite{bernal1933theory}) are allowed, denoted t-I and t-II of energies $\epsilon_{\mathrm{I}}, \epsilon_{\mathrm{I}}$. %,   of energy $\epsilon_{\mathrm{I}}<\epsilon_{\mathrm{II}}$ as in Fig.~1(a). 

The {\it Rys F-model} is a particular six-vertex model whose energies are $0=\epsilon_{\mathrm{I}}<\epsilon_{\mathrm{II}}$. A spin configuration has energy
\begin{equation}
{\cal H}=\epsilon_{\mathrm{II}} N_{\mathrm{II}}
\label{H1}
\end{equation}
 where $N_{\mathrm{II}}$ is the number of t-II vertices. %and all the equilibrium properties can be deduced from the partition function $Z(T)=\sum_{{\cal C}} \exp\left[-{\cal H}({\cal C})/T\right]$, where $T$ is temperature measured as an energy. 

The F-model is invariant under the  $ \mathbb{Z}_2$ time reversal symmetry,  parity symmetry  $A \leftrightarrow B$ (where A, B are the the alternating $A$, $B$ sublattices), and discretized translations.
Its two ordered ground states are  antiferromagnetic tessellations of t-I, have opposite staggered~\cite{baxter1973spontaneous} order parameter $\psi=\pm1$, and thus break the $ \mathbb{Z}_2$ symmetry. Hence, one expects a continuous transition. In fact we know from the exact solution that the transition {\it infinitely} continuous with algebraic correlations for $T>T_c=\epsilon_{\mathrm{II}}/\ln2$. 

In the {\it two-dimensional ice model}, instead, $\epsilon_{\mathrm{I}}= \epsilon_{\mathrm{II}}$, there is no energy scale and thus no transition. The system mimics the degeneracy of water ice in a two-dimensional system and was also solved by Lieb~\cite{lieb1967residual}. It also describes the ground state of degenerate square ice~\cite{perrin2016extensive} and the infinite $T$ limit of the F-model 
   %At infinite temperature the F-model becomes the {\it two-dimensional ice model} (for which $\epsilon_{\mathrm{I}}= \epsilon_{\mathrm{II}}$) also solved by Lieb~\cite{lieb1967residual}, which mimics the degeneracy of water ice~\cite{Pauling1935}. 
%The   model thus realizes long range order below $T_c$ and classical topological order~\cite{henley2011classical,castelnovo2012spin} above $T_c$. % ({\it a latere}: both type of  phases can be seen now in artificial antiferromagnets~\cite{nisoli2018unexpected,libal2018ice,nisoli2017deliberate,Morrison2013,Chern2013,gilbert2014emergent,perrin2016extensive,lao2018classical,drisko2017topological}).

\section{Faraday loops}

The crux of our approach consists in choosing the proper description for the magnetic texture. In the antiferromagnetic ground state, the local magnetization is non-zero, but its coarse graining over a vertex  is zero.  Therefore, instead of considering the elementary spins $\vec S_i$, we describe magnetization by assigning it to   to the vertices $v$, as $\vec M_v$, such that $\sum_i \vec S_i=\sum_v \vec M_v$. 

The t-I vertices are demagnetized, while t-II ones carry magnetization. 
Because of the topological constraints, the magnetic moments of t-II vertices are  always joined into Faraday lines. Thus, Faraday lines carry the magnetization {\it and} energy of the system. Moreover, on a torus, they are always {\em directed} loops and can be distinguished by topological triviality and parity. A combination of parity and topology determines their chirality. Finally, they separate antiferromagnetic domains of opposite staggered order parameter. All of this we show below.

\begin{figure}[t!]
\includegraphics[width=.99\columnwidth]{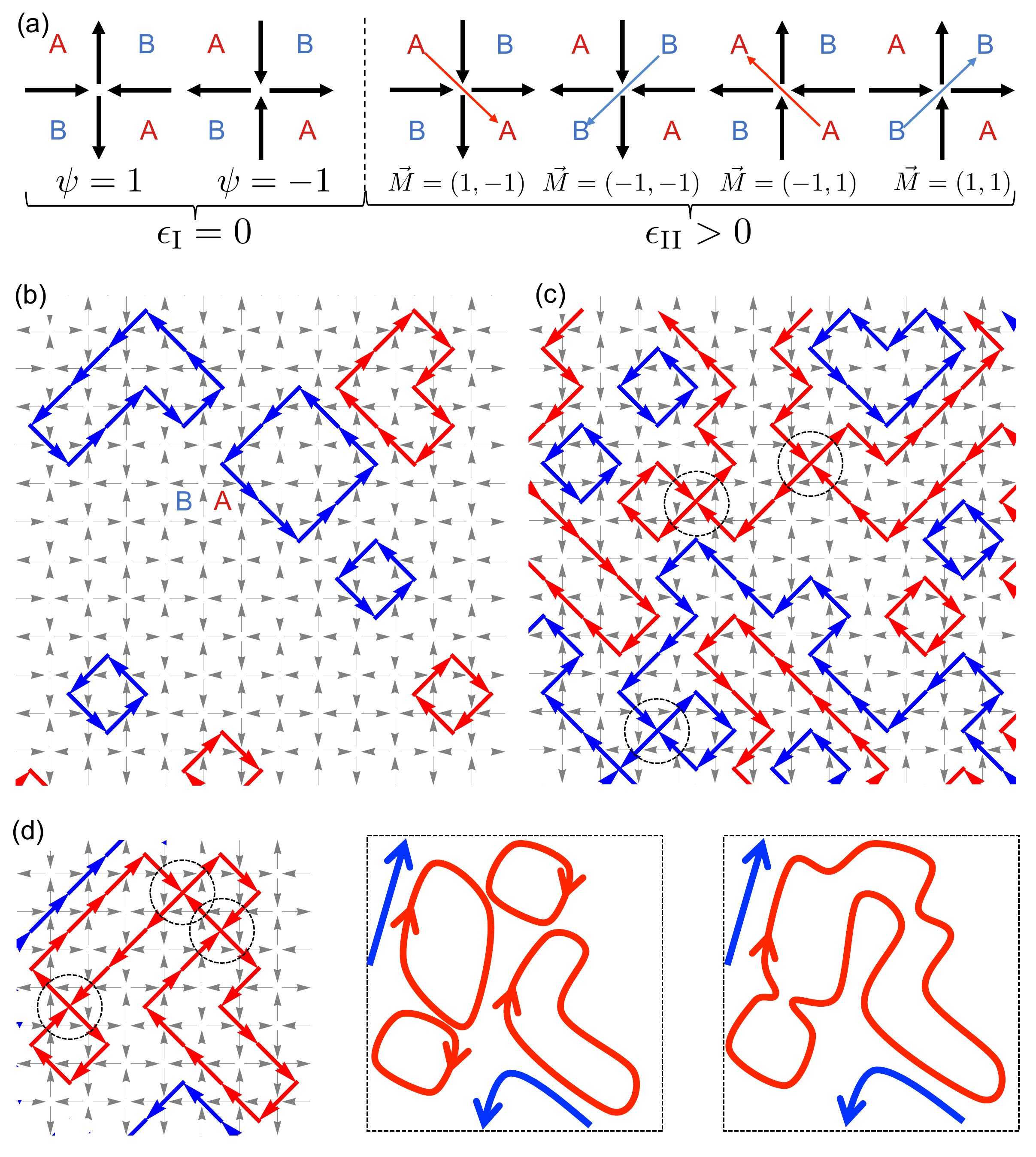}
\caption{(a) The six ice-rule obeying vertices. t-Is  (left of the dotted line) possess a staggered antiferromagnetic order parameter $\psi=\pm1$.  t-IIs carry energy and magnetic moment. Connecting their net moments  (red-blue diagonal arrows for A-B parity of the plaquettes they join), we obtain a directed loop-representation. (b-c) Loops in low  (b) and high (c) energy configurations  have defined parity  and separate antiferromagnetic domains. (d) A spin configuration with  $3$  pinch points and 2 of its  $2^3=8$  loop-representations.}
\label{fig1}
\end{figure}

Consider  $L_x \times L_y$ vertices on a torus ($L_x, L_y$ even). We proceed on a torus for clarity,  and in the thermodynamic limit our considerations are transferrable to the standard 2D system. There, the topological group of the torus corresponds to Faraday lines extending from and to infinity. 

A t-II vertex  can be represented by an arrow connecting the centers of the plaquettes whose spins impinge antiferromagnetically in the vertex, thus assigning to it the magnetization $M_x=\pm1$, $M_y=\pm1$ (Fig.~1a). 

The following  propositions can be verified directly: 

(i) {\em Any spin configuration  can be mapped into a set of non-intersecting, directed Faraday loops}: indeed, a square plaquette can support 0, 2, or 4 t-IIs on its vertices. If 2, they can always be connected unambiguously. If  4, which we call a {\it pinch}, they can be joined in two directed lines in two ways. For a configuration with $P$ pinches, there are $2^P$  loop-representations (Fig.~1b-d). 

(ii) {\em Loops have a  defined parity}:  with the usual alternating $A/B$ assignment of plaquette parity, a loop will only cross either $A$ or $B$ plaquettes. 

(iii)  $A$ and $B$ loops (red and blue  in Fig.~1)  cannot cross. 

(iv) The direction of loops is assigned thus: two nearby loops have the same (resp. opposite) direction if and only if they have same (resp. opposite) parity. If a loop is directly contained into another loop, the two have the same (resp. opposite)  direction if and only if they have opposite (resp. same) parity.  %This univocally assigns the mutual orientation of the loops. %We say that a loop $x$ is {\it directly contained} in another loop $y$ if it is contained in $y$ but is not contained in another loop contained in $y$. 
%  if two topologically trivial loops are not contained in any other loop, or are both {\it directly} contained in the same loop, the two loops have opposite (resp. same) chirality if they have the opposite (resp. same) parity. If a loop is  directly contained into another loop  the two loops have opposite (resp. same) chirality if the have the same (resp. opposite) parity. 

Note (Fig.~1) that loops are also {\it domain walls} separating anti-ferrromagnetic domains with opposite sign of the staggered order parameter $\psi$. For completeness, we show in SI  how the Faraday lines relate to the height formalism.

 Modulo the pinch points, the  spins-loops correspondence is bijective. Any set of directed loops drawn on the square lattice  
such that (i)-(iv) hold true corresponds to an acceptable spin configuration for the six-vertex model.

A trivial loop of the torus is one that can be contracted to zero. We say that a configuration is topologically non-trivial if at least one of his representation contains at least one non-trivial loop  (and then their number  must be even).

Faraday loops  are the elementary excitations of the system, and the F-model is a loop gas. While only (and all) loops carry {\it local} magnetization, crucially, only topologically non-trivial loops carry {\it net} magnetization. 
Indeed, given a loop $\gamma$ made of vertices $v$, the total in-plane magnetization of the  loop is 
\begin{equation}
\vec M_{\gamma}=\sum_{v\in\gamma} \vec M_v. 
\label{M}
\end{equation}
Then $\vec M_{\gamma}=0$ if and only if  $\gamma$ is topologically trivial. If instead $\gamma$  wraps  around the $x$ direction the net magnetization of the loops is $M_y=0$, $M_x=   \pm L_x \hat e_x$
{\em regardless of the length or shape of the loop}. 

We have reached a remarkable result: in the six-vertex model, {\em magnetization is  a homotopy invariant} of the Faraday lines description. This implies that topologically trivial spin configurations have zero net magnetization, and do not couple with an external field. Therefore, the effect of a net Zeeman coupling can only be to induce topological transitions from trivial to non trivial configurations. This can be understood in terms of topological sectors.

%Loops also carry the energy of the system. $A$ configuration $C$  has energy energy 
%
%${\cal H}({\cal C})=\sum_{\gamma \in C} \left( \epsilon_{\mathrm{II}} l_{\gamma} -  {\vec H} \cdot \vec M_{\gamma} \right)$,
 %where $l_{\gamma}$ is the length of  $\gamma$. %Note that the formula is invariant over the possible $2^P$ representations of a configuration. 
 %In SI we show how loops relate to the height function.% Elsewhere we will attempt  an effective gauge theory based on the height function(s).

\section{Topological sectors}

We can now {\em partition} the phase space ${\cal P}$ (the set of all spin configurations) into topological sectors (subsets of defined topology). 

Call ${\cal T}$ the  sector of all  topologically trivial configurations, and  ${\cal W}$ its complementary. From Eq.~(\ref{M}), only configurations in ${\cal W}$ can have magnetization and 
we can further partition it accordingly. 

We call a {\it trivial} (resp. non-trivial) {\it elementary update}  of a spin configuration the flip of %a {\it flippable  loop}, {\it i.e.} 
a trivial (resp. non-trivial) loop of spins that are all head to toe. Consider $n_x$ pairs of alternating $A$ and $B$ non-trivial loops in the direction $x$  (Fig.~2, second row, has $n=2$), with $0< n_x \le L_y/2$.  From (iv),  their magnetization has the same direction. Call  ${\cal M}_{\pm n_x0}$ the set of all topologically trivial updates of such configurations.  

Because of homotopy invariance, trivial updates do not alter magnetization: from Eq.~(\ref{M}), each configuration in  ${\cal M}_{n_x 0}$ carries magnetization $M_y=0$, $M_x=2 n_x L_x$ and magnetization density $m_y=0, m_x=2n_x/L_y$. The same can be done to generate the  sector $ {\cal M}_{0 n_y}$ and for $ {\cal M}_{n_x n_y}$, as the reader can verify via pairs of parallel helices. % of proper pitch.

Crucially, the union  (which we call ${\cal M}$) of these magnetic sectors does not exhaust ${\cal W}$. Call ${\cal W}_0$ the set of non-trivial configurations that have zero net magnetization. %, so that 
 %${\cal W}= {\cal W}_0 \bigcup {\cal M}$. 
 We can  partition ${\cal W}_0$ into: % ${\cal W}_{A_x B_x}$, ${\cal W}_{A_y B_y}$, ${\cal W}_{A_x A_y}$, ${\cal W}_{B_x B_y}$. Fig.~2 shows that 
 ${\cal W}_{A_x B_x}$ (resp. ${\cal W}_{A_y B_y}$),  the sets of all configurations representable via non-trivial loops of type $A$ {\it and} $B$ in the $x$ (resp. $y$) direction;  and ${\cal W}_{A_x A_y}$ (resp. ${\cal W}_{B_x B_y}$) the sets of all configurations  representable via non-trivial loops of parity $A$ (resp. $B$) wrapping in both $x$ and $y$ directions (Fig.~2 bottom). Proposition (iii) forbids other sectors.

In summary, ${\cal P}$ is partitioned into ${\cal T}$ (trivial) and ${\cal W}$ (winding). ${\cal W}$ is partitioned into ${\cal M}$ (magnetic) and  ${\cal W}_0$ (non magnetic). ${\cal W}_0$ is partitioned into ${\cal W}_{A_x A_y}$ (A loop winding in both directions, and same for B)  ${\cal W}_{A_x B_x}$ (A and B loops winding in the x direction, but alternating as to give zero magnetization, and same for $y$).
 
We can introduce winding topological order parameters, $w_A$ and $w_B$, for each parity. For a configuration $C$ and its (possibly multiple) loop-representation(s) $R$ made of loops $\gamma$ we  define
\begin{equation}
w_A(C)=\sup_{R\in C} \sum_{\gamma_A \in R}\left|\frac{1}{L_x L_y}\sum_{v \in \gamma_A}\vec{M_v}\right| 
\label{wr}
\end{equation} 
%
% the number of topologically non-trivial  $A$-loops in the representation $R$ (and  same for B). From this  representation-dependent number we can then define
%%
%\begin{equation}
%w_A=\max_{R\in C} w_A(R)
%\label{w}
%\end{equation} 
%%
the {\it density of winding number} of $A$ loops of the configuration $C$, and  $w^+=w_A+w_B$, $w^-=w_A-w_B$. 

\begin{figure}[t!]
\includegraphics[width=.99\columnwidth]{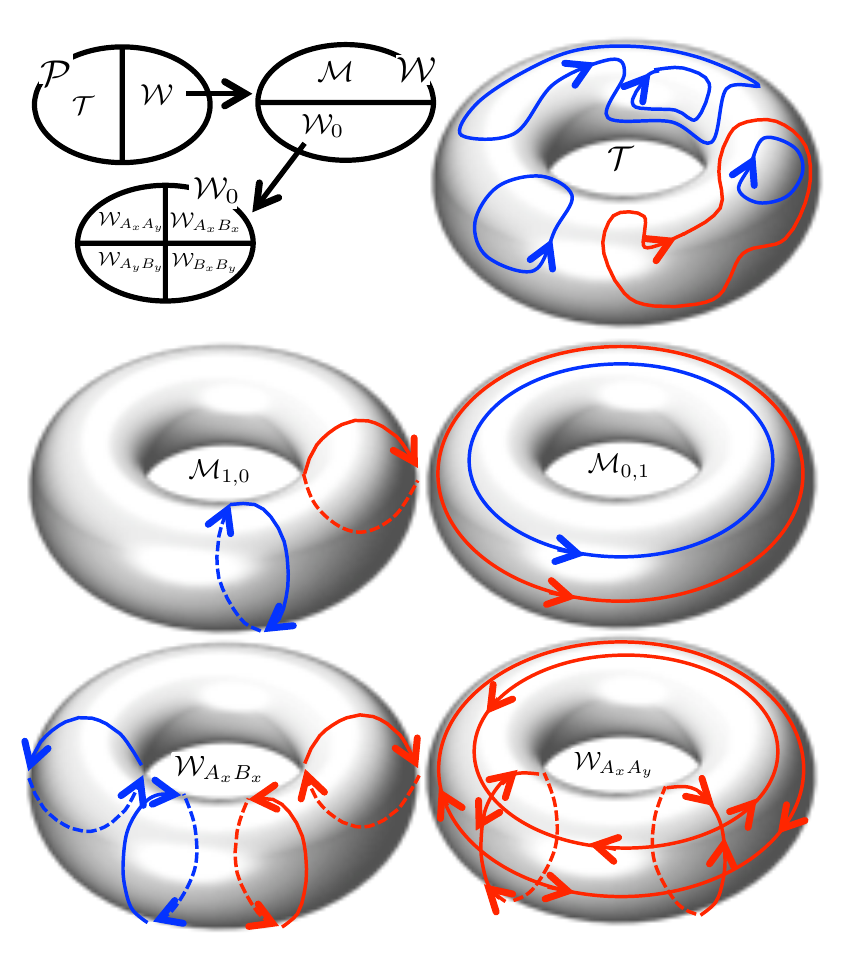}
\caption{Venn diagrams of the partition of the phase space into topological sectors and schematics for elements of ${\cal T}$ (top), ${\cal M}$ (middle), and  ${\cal W}_0$ (bottom) represented on the torus. ${\cal P}$ is partitioned into ${\cal T}$ and ${\cal W}$, corresponding to trivial and non-trivial loops. ${\cal W}$ is partitioned into ${\cal M}$ (corresponding to magnetized configuration) and ${\cal W}_0$.  ${\cal W}_0$ is partitioned into sectors corresponding to loops of different parity.} 
\label{fig1}
\end{figure}

\section{Temperature transition}
When the phase space is partitioned into sectors ${\cal D} \subset {\cal P}$,  we can call %
\begin{equation}
Z_{{\cal D}}(T)=\sum_{{\cal C} \in {\cal D}} \exp\left[-{\cal H}({\cal C})/T\right],
\label{Zd}
\end{equation}
whose sum is restricted to configurations in ${\cal D}$,  the partition function of  ${\cal D}$.  $P_{\cal D}=Z_{\cal D}/Z$ is then the probability of finding the system in a configuration of the sector ${\cal D}$. 
Any observable is said to be {\it limited to the sector} ${\cal D}$ if  obtained from $F_{\cal{D}}=-T \ln Z_{{\cal D}}$. The total partition function is the sum of the partition functions of the sectors.
 
 If $P_{\cal D} \to 1^-$ in the thermodynamic limit (and thus $F\to F_{\cal D}$), we say that the system is {\it asymptotically confined} to  sector ${\cal D}$ and that the partition function {\it  projects} to that sector.  
%When that happens, we call {\it asymptotically irrelevant} any sector in the complementary of ${\cal D}$. 
In this language, a phase transition corresponds to the system ``switching'' between different sectors of the phase space, to which it is confined in the thermodynamic limit. When those sectors are topologically distinct, we say that the transition is topological. 

First we show that in absence of a field, the system is asymptotically confined to   ${\cal T}\cup {\cal W}_0$, that is the complementary of ${\cal M}$. Indeed, consider  $f(m,T)$, the density of free energy limited to the sector ${\cal M}_{m L_y/2,0}$. %corresponding to $n_x =m L_y/2$ non-contractible couples of A-B loops in the $x$ direction. 
Then $ H_x =\partial_{m}f(m,T)$ is the magnetic field.  $f(m)$ must be concave and even in $m$, thus has minimum at $m=0$.  
%\subsubsection{Temperature Transition}

Note that the thermal average of $\psi$  in ${\cal W}$ is zero. To prove it, consider e.g.\ ${\cal M}_{1,0}$.  Its lowest energy state is degenerate, corresponding  to one $A$ and one $B$ non-trivial loops each of length $L_x$, variously assigned, subdividing the torus in two domains of opposite $\psi$. Averaging $\psi$ over all those configurations then returns zero. The same argument can be replicated for any sector in ${\cal W}$. 

That only ${\cal T}$  can exhibit long range order and thus $\psi\ne 0$ only in ${\cal T}$ is intuitive. Indeed, the symmetry breaking that leads to $\psi\ne0$ is driven by the contraction of the domain walls (Faraday loops) as temperature is lowered because of their tensile strength. But outside of  ${\cal T}$, by definition, there are always some non-contractible loops.

Thus, antiferromagnetic ordering in the absence of a field  must correspond to a transition between the topological sectors ${\cal T}$ and ${\cal W}_0$. %and we provide a heuristic argument \`a la Kosterlitz. 
Consider a  non-trivial loop of lowest energy winding around the $x$ axis. It has length  $L_x$  and degeneracy ${{L_x}\choose{L_x/2}}\sim2^{L_x}$, for large $L_x$. Its free energy is then %, as the loop choses a direction in $L_x/2$  of the $L_x$ possible choices. %  implies a free energy $\Delta { F}=  \epsilon_{\mathrm{II}}L_x -T \ln{{L_x}\choose{L_x/2}}$ because the loop choses the downward direction in $L_x/2$  of the $L_x$ possible choices. 
%Then, its free energy  is
%
\begin{equation}
\Delta { F}= L_x (\epsilon_{\mathrm{II}} -T \ln2),
\label{df}
\end{equation}
and goes to $-\infty$ ($ +\infty)$ in the thermodynamic limit for $T>T_c$  ($T<T_c$) with $T_c=\epsilon_{\mathrm{II}}/ \ln2$.  

\begin{figure}[t!]
\includegraphics[width=.99\columnwidth]{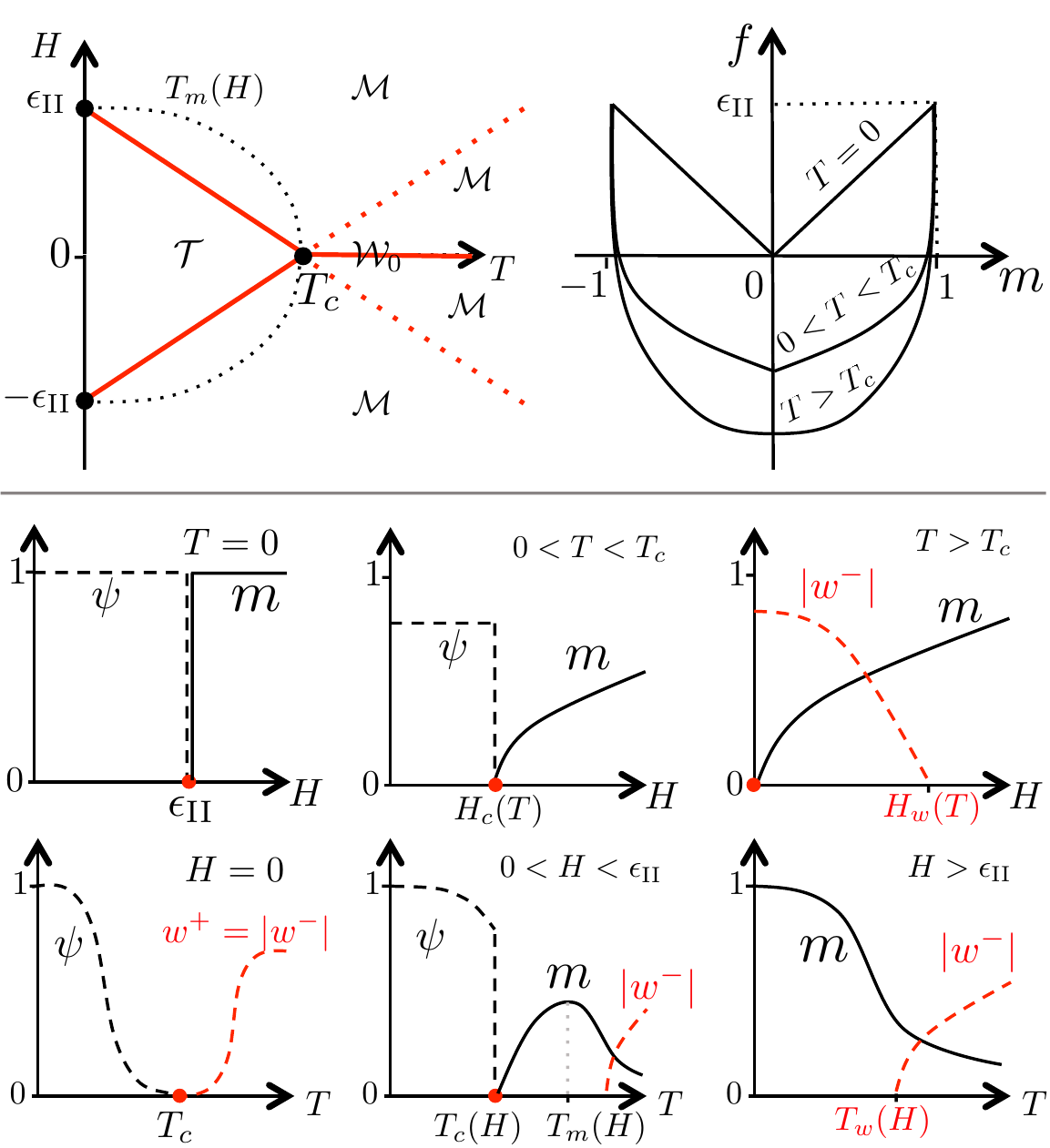}
\caption{First Panel: schematics of the phase diagram  for an horizontal field $H$. The red line marks the distinction between  topological sectors and is critical. On the line $H=0$, $T>T_c$ the system is confined to ${\cal W}_0$, with $m=0$, $\psi=0$, and it is critical. We sketch a possible $T_m(H)$ (black dotted) and the conjectured $T_w(H)$ (red dotted). Second panel: sketches of  $f(m,T)$. The singularity in $m=0$ at $0\le T<T_c$   disappears at $T \ge T_c$. Other panels: sketches  of  $\psi$, $m$, $w^-$ vs.  $T$  and $H$ as deduced from the top two panels and considerations in text.}
\label{fig1}
\end{figure}

As in the heuristic  argument of Kosterlitz and Thouless, Eq.~(\ref{df}) implies that above $T_c$ the system is asymptotically confined to the topologically non-trivial sector ${\cal W}_0$, where $\psi=0$, and  below $T_c$ to the  trivial ${\cal T}$ where $\psi\ne0$.  Therefore, $T_c=\epsilon_{\mathrm{II}}/ \ln2$ is the N\'eel temperature. Crucially, it corresponds to the N\'eel temperature in Lieb's solution.

 Low $T$ configurations correspond to an antiferromagnetic background decorated by Faraday loops (domain walls, Fig.~1b). As $T$  increases, loops grow and coalesce  forming at $T_c$ a topologically non-trivial network (Fig.~1c),  in  a classical analogue to a string-net condensation~\cite{levin2005string}. 

There is more. We have seen that ${\cal T}$  hosts  a  $\mathbb{Z}_2$ symmetry breaking in the sign  of $\psi$.  But  a topological symmetry breaking also exists in ${\cal W}$, between the topological sectors ${\cal W}_{A_x A_y}$ and ${\cal W}_{B_x B_y}$ as they have the same free energy but different parity. At $T>T_c$ the systems must choose whether the network of winding loops has parity $A$ or $B$, because loops of different parity cannot cross. This leads to  a breaking of the $A \leftrightarrow B$ parity symmetry of the topologically non-trivial loops and thus in the sign of $w^-$. 

Thus, at $T<T_c$ we have $w_A=w_B=0$, $\psi>0$. At $T>T_c$ we have $\psi=0$, $w^+> 0$ and $w^-=\pm w^+$. We suspect that $w^+, w^-$ reach zero infinitely continuously as $T\to T_c^+$  just like $\psi$~\cite{baxter1973spontaneous} does for  $T\to T_c^-$, though we are incapable of  predicting it within our framework. 

%We note that while  ${\cal T}$ and ${\cal W}_0$ are topologically distinct,  the loop condensation between ${\cal T}$ and ${\cal W}_0$ proceeds via trivial updates whereas sectors in ${\cal M}$ differ by non-trivial updates, perhaps explaining the discontinuity in $\psi$ in the field . 

\section{Field induced transitions}
We also have transitions under field between demagnetized ${\cal T}$ and magnetized ${\cal M}$, and there  $\psi $ is discontinuous except at $T=T_c$, as we show below.

Consider an horizontal field and let us study the shape of $f(m)$. 
At $T=0$, the free energy is  trivially $f(m)=\epsilon_{\mathrm{II}}m$, and the curve of magnetization is a step function  ($m=0$ for  $H<\epsilon_{\mathrm{II}}$ and  $m=1$ for $H>\epsilon_{\mathrm{II}}$). 

The sector of saturated magnetization  ${\cal W}_{L_y/2,0}$  ($m=1$)    contains configurations all of the same energy, where all the horizontal spins are pointing to the right, whereas half of the vertical rows point up and half down. Its entropy is subextensive [its  degeneracy being  ${{L_y}\choose{L_y/2}}$], and therefore $f(\pm1,T)=\epsilon_{\mathrm{II}}$ at any temperature. 

%Because entropy favors spreading a small number of t-I, and because their excitations to t-II would require formation of new closed loops that cannot be joined, the density of t-II is approximatively $m$  in this limit and the internal energy is  $u(m)\simeq\epsilon_{\mathrm{II}} m$ (at any $T$). If we neglect their correlation when $1-m$ is small, from counting arguments the entropy per vertex $s(m,T)=-\partial_T f(m,T)$ scales as  $s(m)\sim  -(1-m)\ln(1-m)$  for $m \to 1^-$, independent of temperature. 
%From $f(m,T)\simeq \epsilon_{\mathrm{II}} m -Ts(m)$, we have $m(H)\to1^-$ exponentially in  $H \to \infty$.

For $m\simeq 0$, at $0<T<T_c$ entropy favors configurations in which $m L_y/2$ horizontal,  non-trivial loops of alternating parity and of magnetization aligned to the left are maximally spaced at a distance  $1/|m|$,  arbitrarily large. The free energy can thus be approximated by a trivial term from the bulk plus a non-trivial contribution from the loops, or from Eq.~(\ref{df})
\begin{equation}
f(m) \simeq f_{\cal T}+|m| (\epsilon_{\mathrm{II}} - T \ln 2) +o(m) ~~\mathrm{for}~m \to 0.
\label{f_n}
\end{equation}
The weak singularity in $m=0$ implies a critical field %The cusp in $\nu=0$ implies  $m=0$ for $|H| < \pm H_c$, where 
\begin{equation}
|H_c(T)|=\epsilon_{\mathrm{II}} - T \ln 2
\label{H_c}
\end{equation}
below which  the system is asymptotically confined to ${\cal T}$ and there is no magnetization. %Equivalently, there is a field dependent critical temperature $T_c(H)$ obtained by inverting Eq.~(\ref{H_c}) such that the system is confined to ${\cal T}$ for $T<T_c(H)$. 

Instead, no critical field exist when $T>T_c$. In such case, there are topologically non-trivial loops even at $m=0$ and thus no singularity in $f(m)$. Intuitively, the non-contractible loops present above criticality can be biased by weak fields to be of the proper alternation of parity.

 Fig.~3 summarizes our results. The  top panel, left, shows the phase diagram expressed in terms of topological sectors. Top panel, right sketches  $f(m)$ at different temperatures, from which curves for $m$ and $\psi$ can be obtained qualitatively (bottom panels). 
 
 When $0 < H<H_c(T)$, $0 \le T <T_c$   the system is asymptotically confined to ${\cal T}$, and the magnetic field has no effect on the free energy. Thus,  $\psi$  drops discontinuously to zero across the critical line (red) as the system switches to the sectors in ${\cal M}$ and magnetization develops. The entire line $H=0$, $T \ge T_c$ corresponds to the system being confined to ${\cal W}_0$ and is critical (with algebraic correlations~\cite{Baxter1982}).  
%When varying $T$ at $H=0$ the transition is between the topological sectors ${\cal T}$ and ${\cal W}_0$ and is infinitely continuous in $\psi$, as we know from ref~\cite{baxter1973spontaneous} but we cannot deduce from our framework. We note, however, that it corresponds to a string-net condensation that  proceeds through trivial updates connection ${\cal T}$ and ${\cal W}_0$, and contains a symmetry breaking between  ${\cal W}_{A_x A_y}$ and ${\cal W}_{B_x B_y}$ in ${\cal W}_0$.      

When  $0<H<\epsilon_{\mathrm{II}}$,  the magnetic moment is non-monotonic in temperature (Fig.~3) because as $T$ increases and crosses the critical line, magnetization ensues, yet  for large $T$  it must tend to zero. We call $T_m=T_m(H)$ the temperature at which the maximum of $m$ is achieved. 
%In the phase diagram, that curve is defined implicitly by
%%
%\begin{equation}
% \partial_ms(m(H,T_{m}),T)=0,
% \label{Tm}
%\end{equation}
%%
We show in SI that  $T_m\to 0$ when  $H\to\epsilon_{\mathrm{II}}^-$, while $T_m\to T_c^-$ when $H\to 0$. %Indeed when $T>T_c$ increasing $m$ implies imposing alternating order on the non-trivial loops, and thus  $s(m,T)$ is aways monotonically decreasing in $m$  and its derivative is never zero. Thus $T_c(H)<T_m(H)<T_c$ and consequently $T_m\to T_c$ when $H\to 0$. If $H>\epsilon_{\mathrm{II}}$ and $T=0$ then $m=1$ which is maximum and thus $T_m(H)=0$ for $H \ge \epsilon_{\mathrm{II}}$. %Furthermore $dT_m/dH\to -\infty$ as $H\to\epsilon_{\mathrm{II}}^-$. Indeed by differentiating Eq.~(\ref{Tm}) we have
%%
%\begin{equation}
% \chi\partial^2_{m}s \frac{dH}{dT_m}+\partial^2_{m,T}s=0.
%\label{dtdh}
%\end{equation}
%%
%As $H \to \epsilon_{\mathrm{II}}^-$ and $T \to 0$, both $\chi$ and, from Eq.~(\ref{s(1-m)}), $\partial^2_{m}s $ diverge while  $\partial^2_{m,T}s=\partial_m c_v/T$ does not ($c_v(m,T)$ is the specific heat per vertex of the sector corresponding to $m$), thus implying divergence in ${dT}/{dH_m}$. As $H \to 0$ and $T \to T_c$, $\chi$ and $\chi\partial^2_{m}s$ tendo to zero while while  $\partial^2_{m,T}s$ remains finite, implying ${dT}/{dH_m} \to 0$. 
A plausible sketch of $T_m(H)$ is shown in Fig.~3.

 Configurations in ${\cal T}_0$ always have $w^+=w^-=0$. Configurations in ${\cal W}_0$ always have $w^+>0$, while $w^-$ can be zero, e.g.\ in ${\cal W}_{A_x B_x}$. It is always $|w^-|=w^+$ (resp $w^-=-w^+$) in  ${\cal W}_{A_x A_y}$ (resp. ${\cal W}_{B_x B_y})$. %We have seen that above $T_c$ at zero field $w^-=\pm w^+>0$. 
For a configuration  with density of magnetization $m$, we have $w^+ \ge m$. Instead, $w^-$ can be zero, and is indeed zero at minimal energy where non-trivial loops perfectly alternate parity. Configurations of higher energy can have $w^->0$  when loops  point in the direction opposite to the magnetization, breaking the parity alternation (and thus $w^+>m$). We conjecture that such loops are possible only when their free energy (inclusive of a Zeeman term) is negative, or $\epsilon_{\mathrm{II}}+|H|-T\ln2<0$, leading to a new  line %$T^w(H)$ with 
%
%\begin{equation}
$
T_w(H)=( \epsilon_{\mathrm{II}}+|H|)/\ln2, 
$
%\label{e+H}
 %\end{equation}
%
for the appearance of the winding order parameter $w^-$. %, which  breaks the symmetry between $A$ and $B$ lines when $T>T^w(H)$ (dotted red line in Fig.~3). 
In Fig.~3 we have sketched possible curves for $w^-$ (red dashed line).

 \begin{figure}[t!!!!!!]
\includegraphics[width=.999\columnwidth]{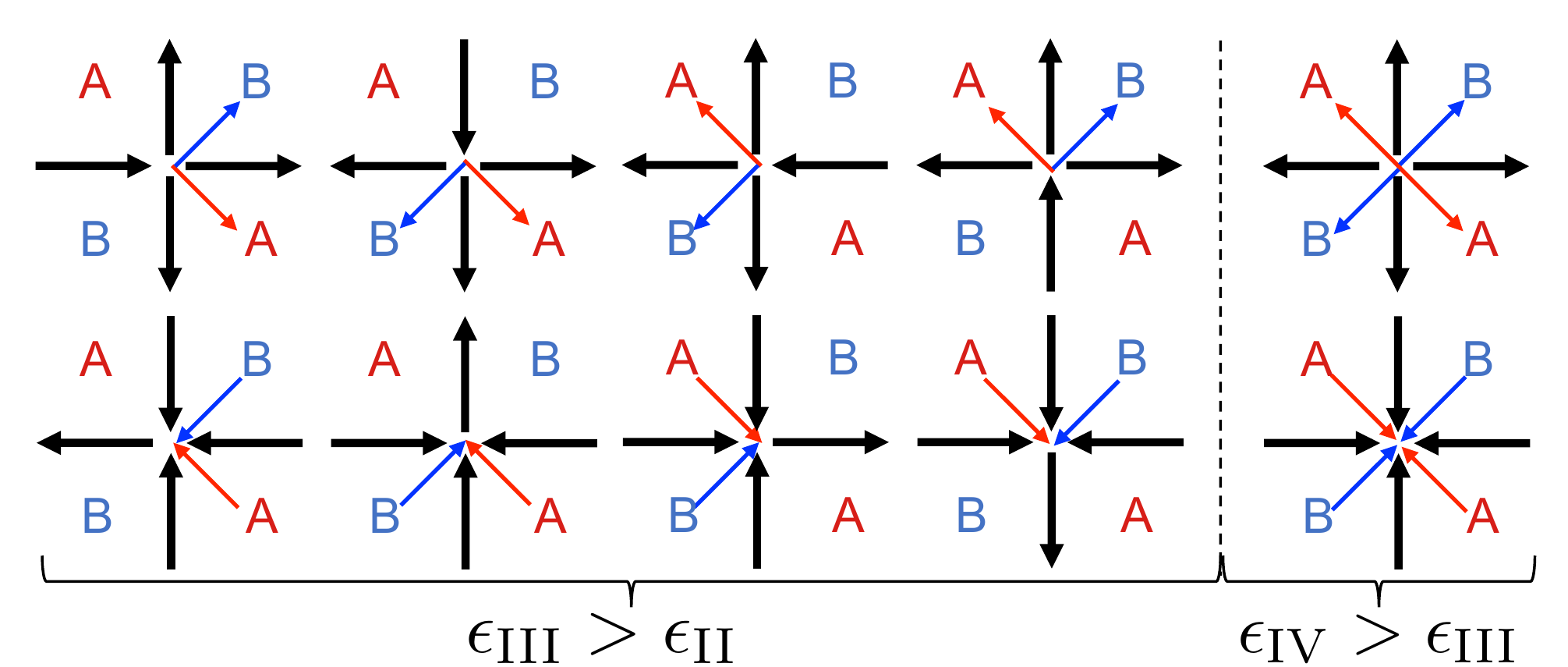}\vspace{3mm}
\includegraphics[width=.47\columnwidth]{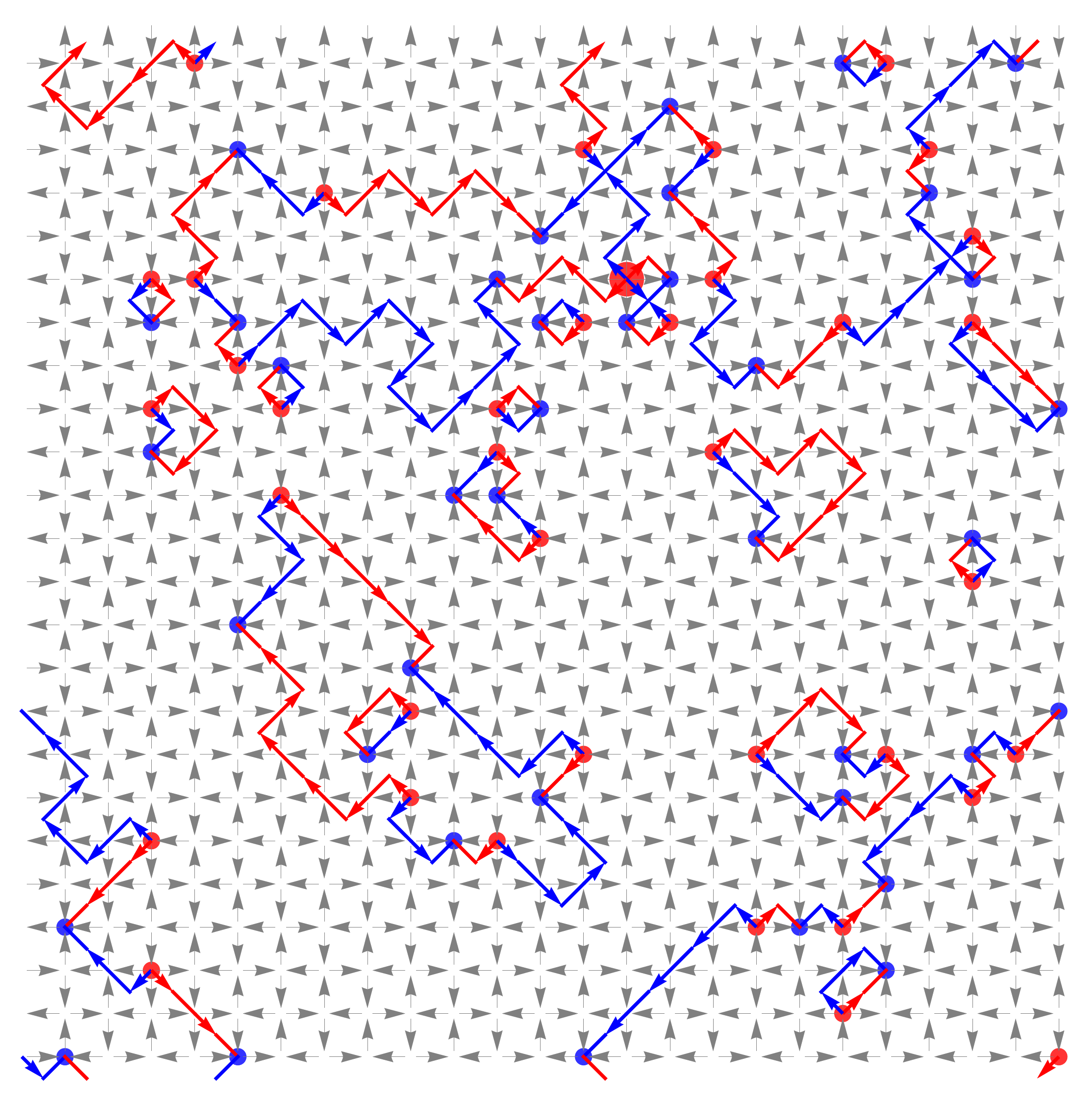}\hspace{4 mm}\includegraphics[width=.47\columnwidth]{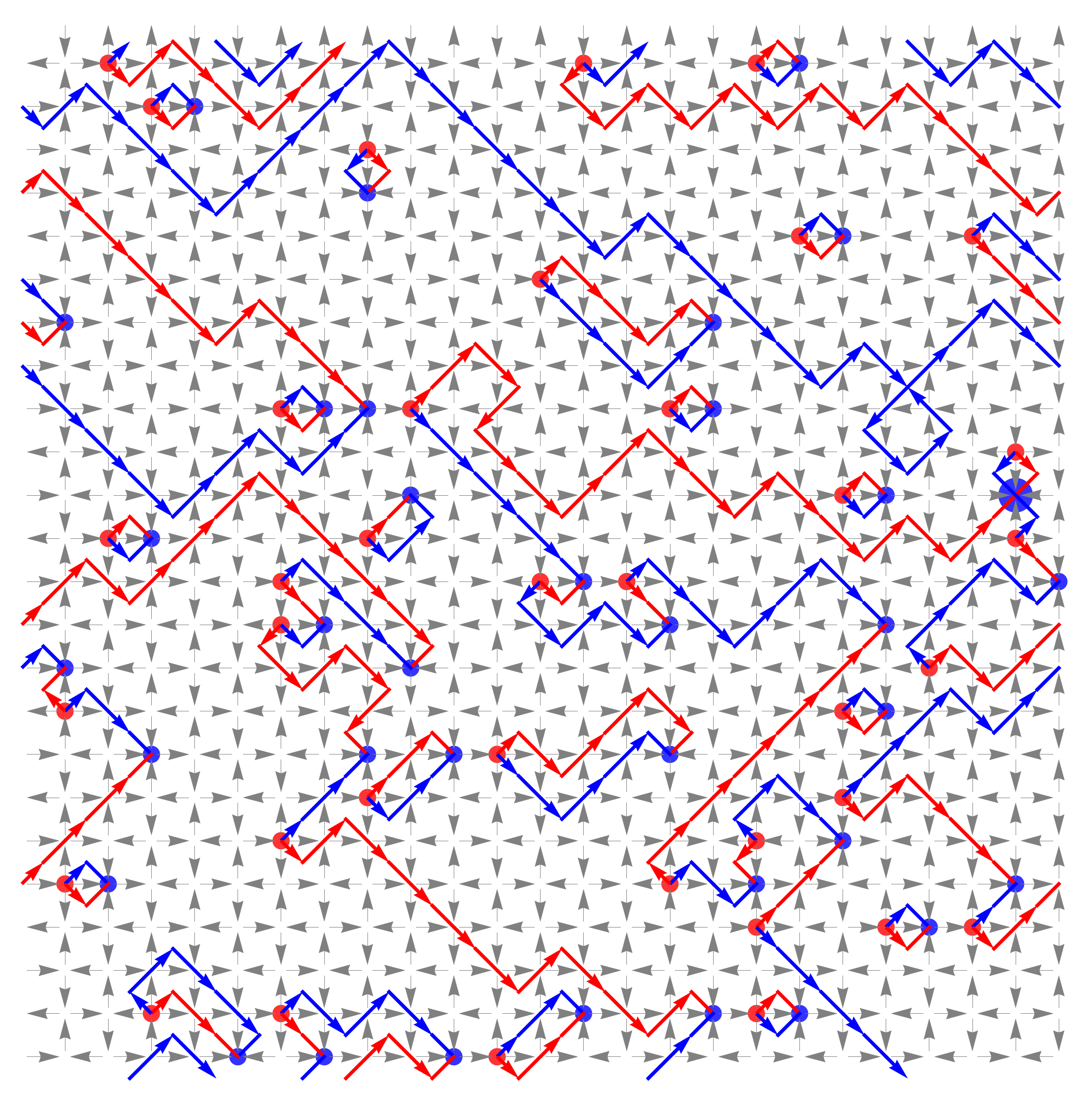}
\caption{Top: The ten extra vertices (monopoles) included in the sixteen-vertex model can also  be represented by arrows separating frustrated spins and whose sum represents the total magnetization. The  t-III (left of dotted line) have topological charges $\pm 2$, magnetizations $\vec {M} =(\pm1,0), (0,\pm1)$, and  energy $\epsilon_{\mathrm{III}}>\epsilon_{\mathrm{II}}$. t-IV vertices have topological charges $\pm 4$, zero magnetization, and energy  $\epsilon_{\mathrm{IV}}>\epsilon_{\mathrm{III}}$. For all, $\psi=0$. In the resulting loops representation (bottom left), domain wall-loops  mix parity and invert magnetization at the monopoles, (red and blue dots for negative and positive respectively). Hence, topologically trivial loops can carry net magnetization (Bottom right shows horizontal magnetization).}
\label{16v}
\end{figure}

\section{Monopoles and Faraday lines in  spin ice}

 Any kinetics of the F-model must involve topologically-trivial updates only, at least in the thermodynamic limit. But sectors in ${\cal M}$ differ by non-trivial updates. This assures  ergodicity breaking~\cite{palmer1982broken}. We find therefore the unphysical conclusion that the system will persist in a magnetized, high energy state forever after the field is removed. Also, its magnetization will be independent on changes of temperature (though its free energy would change).  
 
This indeed does not  happen in real systems such as square spin ices~\cite{Ramirez1999,Wang2006,Nisoli2013colloquium}, which instead evolve via individual spin flips, which are forbidden in the F-model as they lead to ice-rule violations, or monopoles.

 At nearest neighbors, square spin ice is described by the {\it sixteen-vertex} model~\cite{Wu1969}, which contains all  the possible vertices. Figure~4 shows  the ten extra vertices, as t-III and t-IV ice rule violating vertices of  energies $\epsilon_{\mathrm{II}}<\epsilon_\mathrm{III}<\epsilon_\mathrm{IV}$, endowed with a topological charge ($\pm 2, \pm4 $) defined as the difference of spins pointing in and out. This energy hierarchy describes the most common magnetic realizations~\cite{Wang2006,Morgan2010,Nisoli2010,levis2013thermal,Porro2013,Zhang2013,cugliandolo2017artificial,sendetskyi2019continuous} and also particle-based ices~\cite{Libal2006,ortiz2016engineering,libal2017,libal2018ice} via a proper mapping at equilibrium~\cite{nisoli2014dumping,nisoli2018unexpected}, though even different hierarchies can be obtained through various clever methods~\cite{Moller2009,perrin2016extensive,farhan2019emergent,ostman2018interaction}.

 Figure~4 shows that monopoles too can be incorporated into a Faraday picture, but modify it essentially. Now monopoles allow  for the mixing of $A$ and $B$ lines, and thus there is no longer parity distinction for the loops. Also, we see that Faraday lines can now be just {\it lines} and not necessarily loops, and monopoles are their sinks and sources. This is the a geometric version of the Gauss' law.  
 
Faraday lines still compose into domain wall loops, These can contain an even number of $\pm2$ monopoles, whose charge  alternates in sign along the loop. Therefore,  magnetization is no longer an homotopy invariant of non-contractible loops. Topologically trivial loops can carry net magnetization if they contain monopoles.  %t-IV monopoles further allow for crossing of $A$ and $B$ lines. 
 
Therefore, the previous partition of the phase space in topological sectors  breaks down. And
  indeed, the  antiferromagnetic transition in square ice is known to be of second order, as the system can be mapped into a frustrated  $J_1$-$J_2$ Ising model~\cite{Wu1969}. Also, by allowing for more entropy to destroy the ordered state,  monopoles lower the critical temperature, or $T_c<\epsilon_{\mathrm{II}}/\ln2$.

We can also understand in this picture why there is not any critical field for magnetization. The domain wall themselves, if they contain monopoles, are magnetizable. Because the magnetization flips verse at the monopole, a trivial loop can carry net magnetization. An external field can align the magnetization of a trivial loop by reallocating its monopoles along the domain wall, and thus the system is magnetically susceptible at any field even in the antiferromagnetic phase. There will still be a critical field $H_c(T)$ for the disappearance of $\psi$  [and clearly $H_c(0)=\epsilon_{\mathrm{II}}$, $H_c(T_c)=0$] but the system can be magnetized  for $H< H_c(T)$ via the paramagnetism of the domain wall loops (Fig.~5),  proportional to the density of t-II. Thus magnetization should scale as $m\sim H\exp(-\epsilon_{\mathrm{II}}/T)/T$. 

%As $H$ increases, the Zeeman energy lowers the energy cost of the two (out of four) t-II configurations whose moment is aligned along the field, thus proliferating field-alligned domain walls that eventually merge. In particular, at $H=\epsilon_{\mathrm{II}}$ the ground state is degenerate, and $m$ is still a step function in $H$ at $T=0$. At $H_c$ the loops condense on a network where monopoles have high mobility, corresponding to a maximum in the magnetic susceptibility, as in Fig.~5. 

Finally, the kinetics of real spin ice loses the topological ergodicity breaking of the Rys F-model precisely because any kinetics  is monopole kinetics: a single spin flip corresponds to either creation-annihilation of a monopole pair or to its motion. In turn, this implies to the nucleation of loops and also their growth, contraction,  or fluctuation, as it will be shown in a future work.

\section{Faraday Lines and Dirac strings}
This representation is useful also when the square ice is degenerate~\cite{Moller2009,perrin2016extensive,farhan2019emergent,ostman2018interaction}, or $\epsilon_{I}=\epsilon_{\mathrm{II}}<\epsilon_{\mathrm{III}}<\epsilon_{\mathrm{IV}}$. There, the ground state is described by the {\it two-dimensional ice model} explained above: monopoles are absent, the topological structure described above is valid, and the system is asymptottically confined to the sector ${\cal W}_0$ at zero applied field and to ${\cal M} $ at any non zero field. It has therefore non-zero susceptibility at any field. 

However, at any  non-zero temperature monopole forms as sources and sinks of the Faraday lines. Therefore the $T=0$ case cannot  be considered a $T\downarrow 0$ limit for degenerate square spin ice, as it is essentially topologically different. In degenerate square spin ice, $T=0$ is therefore the essentially non-perturbative critical point for a topological phase, thus explaining why the correlation length $\xi$ tends to infinity faster than algebraically~\cite{nisoli2020field,fennell2009magnetic}. Moreover, in physical realizations the entire model generally breaks down at low temperature in real systems.

In future work, we will treat more in depth the the relationship between Faraday lines, the so-called Dirac strings, the gauge freedom associated to the height function representation, and magnetic fragmentation~\cite{brooks2014magnetic,canals2016fragmentation,petit2016observation}.

Here we note that for an antiferromagnetic square ice the language of Dirac strings is often unsuitable. 
It is often said that monopoles in square ice are ``linearly confined'' by Dirac strings. That is in general not true. 

Looking at Fig~4 the reader can verify that two monopoles connected by Faraday lines can feel no force bringing them together---or indeed no force at all: it is generally {\em not true} that a spin always exists, impinging on a given monopole, whose flipping reduces the energy by moving the monopole. Certainly, Faraday lines have tensile strength, but that is true generally in the F-model where monopoles are absent. 

We can talk about Dirac strings of tensile strength merely in the case in which a monopole pair can be annihilated on a  t-I antiferromagnetic tessellation by flipping {\em a single directed line of spins}. While interesting, that does not describe, however, all the possible situations. Furthermore, such a case is still better described by two Faraday lines running parallel to the claimed Dirac string, both starting from the negative monopoles and ending in the positive one (many examples are visible in Fig.~4), as Faraday lines unequivocally carry   energy and magnetization. 

\begin{figure}[t!!!!!!]
\includegraphics[width=.99\columnwidth]{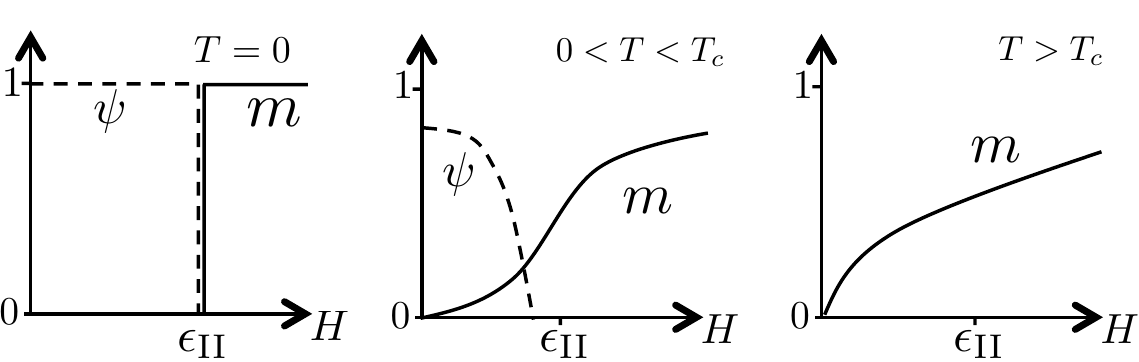}
\caption{Sketches of $m$ and $\psi$  in a square ice. At $T=0$ we have the usual step function as a pointwise limit. The case below $T_c$ differs considerably from the F-model. There is small but non-zero susceptibility at low field, and the maximum susceptibility corresponds to a left neighborhood of $\epsilon_{\mathrm{II}}$ where $\psi=0$.  }
\label{Mag_ASI}
\end{figure}

\section{Conclusion}
We have introduced a Faraday Line picture for square ice in general, and the six-vertex model in particular. It allows for a partition of the phase space into topological sections, shedding light on the topological nature of the transitions of the Rys F-model that. In spin ice materials this representation survives, but  its topological features break down because of monopoles and the Gauss' law. Within this picture, it would be interesting to investigate scaling limits in which the transition in spin ice tends to the topological.  Furthermore, our picture of topological non-trivial loops can be employed in finite-size realizations of artificial spin ice on a cylinder, which are now possible~\cite{gliga2019architectural}, or in attacking  problems of the six-vertex model with fixed boundary conditions~\cite{Baxter1982,bogoliubov2002boundary, pannevis2012critical,zinn2000six,zinn2009six,barkema1998monte,weigel2005square,keesman2017numerical,eloranta1999diamond,van1977exactly}.

\acknowledgments
We  thank C. Castelnovo (Cambridge) and C. Batista (Tennessee) for useful feedback. This work was carried out under the auspices of the U.S.
DoE through the Los Alamos National
Laboratory, operated by
Triad National Security, LLC
(Contract No. 892333218NCA000001).

\bibliographystyle{eplbib.bst}

\bibliography{library2.bib}{}

\end{document}